\documentclass[12pt,twoside,a4paper]{article}
\usepackage{amsmath,amsfonts,amssymb}
\scrollmode

\def\d{{\rm d}}
\def\Lie{{\pounds}}

\def\half{{\textstyle\frac{1}{2}}}

\def\Preprint#1{\vbox to 0pt{\vskip-20mm\rightline{\normalsize\rm#1}\vss}}

\begin{document}

\title{
\Preprint{ADP-98-38/M69}
\bf On Waylen's regular axisymmetric similarity solutions
}

\author{
Edward D Fackerell\footnote{
School of Mathematics and Statistics F07,
University of Sydney, Australia 2006}
\and
David Hartley\footnote{
Physics and Mathematical Physics,
University of Adelaide, Australia 5005}
}

\date{30th September 1998}

\maketitle

\begin{abstract}
We review the similarity solutions proposed by Waylen for a regular
time-dependent axisymmetric vacuum space-time, and show that the key
equation introduced to solve the invariant surface conditions is related by
a B{\"a}cklund transform to a restriction on the similarity variables. We
further show that the vacuum space-times produced via this path
automatically possess a (possibly homothetic) Killing vector, which may be
time-like.
\end{abstract}




\section{Introduction}

Waylen \cite{W87} introduced a general solution for a time-dependent,
axisymmetric, vacuum gravitational field, taking the form of a convergent
radial power series whose coefficients are determined by a single arbitrary
generating function with cylindrical coordinate expression
$a(t,z)$. Motivated by this work, Waylen \cite{W93} then examined the
problem of obtaining exact forms of these series solutions, and found that
for similarity solutions, the invariants could be written in
closed form in terms of a so-called {\em key function\/} $\chi$. A
sufficient condition for the existence of such a solution is a third-order
quasi-linear partial differential equation for $\chi$, here called {\em
Waylen's equation}:
\begin{multline}\label{waylen}
	(\chi_t - 2Kz)(\chi_{tt}\chi_{zzz} - \chi_{zz}\chi_{ttz})\\
	+ (\chi_z + 2Kt)(\chi_{zz}\chi_{ttt} - \chi_{tt}\chi_{tzz})
	+ 4\chi_{tt}\chi_{tz}\chi_{zz} = 0,
\end{multline}
where $K$ is a constant parameter.

Given a solution $\chi(t,z)$ to equation \eqref{waylen}, the similarity
variables and forms are then known explicitly. Substituting these into the
vacuum field equations yields a reduced set of partial differential
equations (whose consistency is guaranteed by equation
\eqref{waylen}). These must be solved before the final metric can be 
written down. Further details of this process can be found in the original
papers already cited.

The aim of this note is to study the structure of Waylen's procedure,
focussing on the point symmetries in the similarity solution and the
connection with Waylen's equation.

\section{Field equations}

In this section, we describe Waylen's reduced field equations and exact
similarity reductions for Einstein's vacuum equations with axisymmetry,
that is, metrics for which there is a single hypersurface-orthogonal
Killing vector $\partial_\phi$.

In the 1987 paper \cite{W87}, Waylen showed that the metric for this case
could be chosen in the form
\begin{equation}\label{metric}
	\d s^2 = A\,\d t^2 +2B\,\d t\,\d z - C\,\d z^2 
		- (AC+B^2)\,\d\rho^2 - \rho^2\,\d\phi^2,
\end{equation}
where $A$, $B$ and $C$ depend upon $\rho$, $t$ and $z$ only. For this
metric, the fundamental equations to be solved were shown to be
\begin{equation}\label{main}
\begin{aligned}
	A_{\rho\rho} + \rho^{-1}\,A_{\rho} 
	&= A_{\rho}E_{\rho} + AH + DE_{tt} + A_{z}B_{t} - A_{t}B_{z},\\
	B_{\rho\rho} + \rho^{-1}\,B_{\rho} 
	&= B_{\rho}E_{\rho} + BH + DE_{tz} + \half(A_{t}C_{z} - A_{z}C_{t}),\\
	C_{\rho\rho} + \rho^{-1}\,C_{\rho} 
	&= C_{\rho}E_{\rho} + CH -  DE_{zz} + C_{t}B_{z} - C_{z}B_{t},
\end{aligned}
\end{equation}
in which the abbreviated expressions are
\begin{equation}
\begin{gathered}
	D = AC + B^2, \qquad E = \log D,\qquad\text{and}\\
	H = C_{tt} + 2B_{tz} - A_{zz} - D^{-1}(A_{\rho}C_{\rho} + B_{\rho}^2).
\end{gathered}
\end{equation}
The remaining components of Einstein's equations for solutions analytic on
the axis of symmetry were satisfied by imposing the boundary conditions
\begin{equation}\label{boundary}
	ac = e^{2k},\qquad b = 0,
\end{equation}
on the axis of symmetry, where $a=A|_{\rho=0}$, $b=B|_{\rho=0}$,
$c=C|_{\rho=0}$ and $k$ is a constant. Using this information, a convergent
series expansion of the general solution was given in terms of $a(t,z)$ and
$k$.

In his 1993 paper \cite{W93}, Waylen stated without proof that the exact
forms of $A$, $B$ and $C$ for similarity solutions of equations
\eqref{main} and \eqref{boundary} are given by
\begin{equation}\label{exactform}
\begin{aligned}
	A &= a\{L + (1 - f^2)^{-1}[2fM + (1+f^2)N]\},\\
	C &= a^{-1}\{L - (1 - f^2)^{-1}[2fM + (1+f^2)N]\},\\
	B &= (1-f^2)^{-1}[ (1+f^2)M + 2fN].
\end{aligned}
\end{equation}
The symmetry invariants $L$, $M$ and $N$ are functions of the two
similarity variables $u = U(t,z)$ and $v = \rho^2V(t,z)$, with
\begin{align}
	U(t,z) &= [a^{-1}(\chi_{t} - 2Kz)]_{t} - [a(\chi_{z} + 2Kt)]_z\\
	V(t,z) &= (a^{-1})_{tt} - a_{zz}.
\end{align}
Here $\chi$ is the key function, and $K$ is a dimensionless
constant. The remaining element $f(t,z)$ of the exact form
\eqref{exactform} is given by
\begin{equation}
	f = a(\chi_{z} + 2Kt)/(\chi_{t} - 2Kz),
\end{equation}
while the boundary data $a(t,z)$ is related to $\chi(t,z)$ through
\begin{equation}\label{a}
	a = \left(-\chi_{tt}/\chi_{zz}\right)^{1/2}.
\end{equation}

Inserting the exact form \eqref{exactform} into the field equations
\eqref{main} yields reduced equations for $L$, $M$ and $N$, whose
integrability demands that $\chi$ satisfy Waylen's equation \eqref{waylen}.

In view of the rich structure involved in the procedure outlined above, it
seems worthwhile to investigate the main field equations \eqref{main} for
Lie point symmetries in order to shed some light on the form of the
solution described.

\section{Point Symmetries}\label{symmetries}

The purpose of this section is to analyze the point symmetries of the field
equations \eqref{main}, paying special attention to the boundary conditions
\eqref{boundary}, and to derive the conditions for the similarity
variables. We also examine the implications for further space-time
symmetries.

Using Langton's enhanced version \cite{L97} of Kersten's computer algebra
package \cite{K87}, we find the point symmetries of the main field
equations \eqref{main} to be generated by the vector field
\begin{equation}
\begin{aligned}
	X = 	& \phantom{+} (c_1\rho + c_2\rho\log\rho)\partial_\rho 
		  + (\chi_z + 2Kt)\partial_t
		  - (\chi_t - 2Kz)\partial_z \\
		& + 2\big([2K - c_1 - c_2(1+\log\rho) - \chi_{tz}]A 
		  	+ \chi_{tt}B\big)\partial_A\\
		& - \big(\chi_{zz}A - 2[2K - c_1 - c_2(1+\log\rho)]B
		  	+ \chi_{tt}C\big)\partial_B\\
		& + 2\big(\chi_{zz}B + [2K - c_1 - c_2(1+\log\rho) 
			- \chi_{tz}]C\big)\partial_C,
\end{aligned}
\end{equation}
where the constants $c_1$, $c_2$ and $K$ take any values, and $\chi(t,z)$
is an arbitrary expression. At this stage, $\chi$ is not restricted to
satisfy Waylen's equation \eqref{waylen}. The point symmetries thus form an
infinite-dimensional Lie pseudogroup.

For solutions analytic on the axis of symmetry $\rho=0$, we remove the
logarithmic terms by setting $c_2=0$, leaving
\begin{equation}\label{symvec1}
\begin{aligned}
	X = 	& \phantom{+} c_1\rho\partial_\rho 
		  + (\chi_z + 2Kt)\partial_t
		  - (\chi_t - 2Kz)\partial_z\\
		& + 2[2K - c_1]\big(A\partial_A + B\partial_B 
			+ C\partial_C\big)\\
		& - 2\big(\chi_{tz}A 
		  	- \chi_{tt}B\big)\partial_A
		  - \big(\chi_{zz}A
		  	+ \chi_{tt}C\big)\partial_B
		  + 2\big(\chi_{zz}B - \chi_{tz}C\big)\partial_C.
\end{aligned}
\end{equation}

The vector \eqref{symvec1} generates symmetries of the main field equations
\eqref{main} alone. These equations are not sufficient by themselves to
guarantee a solution of Einstein's vacuum equations. We must further ensure
that the boundary conditions \eqref{boundary} are also preserved under the
symmetry transformations, up to a possible change in the constant
$k$. Rewriting the boundary conditions as the simultaneous equations
\begin{equation}
	\rho = A_t C + A C_t = A_z C + A C_z = B = 0,
\end{equation}
they will be preserved provided the conditions
\begin{gather}
	c_1 = 2K\label{constants}\\
	\chi_{tt} + a^2\chi_{zz} = 0\label{chi2}
\end{gather}
are satisfied. The first condition \eqref{constants} on the constants will
re-appear below, while the constraint \eqref{chi2} on the previously
arbitrary function $\chi$ is the origin of the relation \eqref{a}.

Leaving aside condition \eqref{constants} for a moment, it is interesting
to ask whether the symmetry vector \eqref{symvec1} of the main field
equations \eqref{main} induces a symmetry of the metric \eqref{metric}
itself. To this end we compute
\begin{equation}
	\Lie_X g = (X g_{\mu\nu})\d x^\mu\otimes\d x^\nu
		+ g_{\mu\nu}\d(X x^\mu)\otimes\d x^\nu
		+ g_{\mu\nu}\d x^\mu\otimes\d(X x^\nu),
\end{equation}
and find for $g$ given by \eqref{metric} that
\begin{equation}
	\Lie_X g = 2(4K - c_1) g + 4(2K - c_1)\rho^2\d\phi\otimes\d\phi.
\end{equation}
Hence condition \eqref{constants}, necessary for the full Ricci tensor to
be preserved, also guarantees that the similarity solution generated by $X$
will have a homothetic Killing vector.

To summarise the considerations so far, point symmetries of the main field
equations \eqref{main} which preserve the boundary conditions
\eqref{boundary} and the metric's analyticity on the axis are generated by
the vector field
\begin{equation}\label{symvec}
\begin{aligned}
	X = 	& \phantom{+} 2K\rho\partial_\rho 
		  + (\chi_z + 2Kt)\partial_t
		  - (\chi_t - 2Kz)\partial_z\\
		& - 2\big(\chi_{tz}A 
		  	- \chi_{tt}B\big)\partial_A
		  - \big(\chi_{zz}A
		  	+ \chi_{tt}C\big)\partial_B
		  + 2\big(\chi_{zz}B - \chi_{tz}C\big)\partial_C,
\end{aligned}
\end{equation}
and result in an axisymmetric space-time with an additional homothetic
Killing vector obeying
\begin{equation}
	\Lie_X g = 4K g.
\end{equation}

Returning to the similarity reduction of the main field equations
\eqref{main}, we seek functions of the variables $t$, $z$, $\rho$, $A$, $B$
and $C$ which are invariants of the symmetry transformation and are thus
annihilated by the vector field $X$. Since the coefficients of
$\partial_t$, $\partial_z$ and $\partial_\rho$ are functions of the
independent variables only, it follows that there are two similarity
variables $u$ and $v$, which may furthermore be taken in the form
$u(t,z,\rho)=U(t,z)$ and $v(t,z,\rho)=\rho^2 V(t,z)$. The conditions $Xu =
Xv = 0$ for $u$ and $v$ to be similarity variables become
\begin{equation}\label{simvar}
\begin{aligned}
	(\chi_z + 2Kt)U_t - (\chi_t - 2Kz)U_z &= 0\\
	4KV + (\chi_z + 2Kt)V_t - (\chi_t - 2Kz)V_z &= 0.
\end{aligned}
\end{equation}
For $U$ and $V$ independent, these can be resolved to give
\begin{equation}\label{simvar1}
	\chi_t - 2Kz = \frac{4KV U_t}{U_t V_z - U_z V_t},\qquad
	\chi_z + 2Kt = \frac{4KV U_z}{U_t V_z - U_z V_t}.
\end{equation}

\section{B{\"a}cklund transform}

Knowing the point symmetries \eqref{symvec} of the field equations and the
invariance conditions \eqref{simvar} for similarity variables, we are still
a long way from Waylen's equation and the exact form \eqref{exactform} of
the similarity variables. In this section we present the link between the
invariance conditions and Waylen's equation as a B{\"a}cklund map, showing that
the latter represents a restriction on the set of possible similarity
solutions.

We start from Waylen's equation \eqref{waylen}, and re-introduce the series
generator $a(t,z)$ given by equation \eqref{a}. Treating $\chi$ and $a$ as
independent, Waylen's equation becomes a quasilinear system of mixed first
and second order
\begin{equation}\label{mixed}
\begin{gathered}
	\chi_{tt} + a^2\chi_{zz} = 0\\
	(\chi_t - 2Kz)a_z - (\chi_z + 2Kt)a_t = 2a\chi_{tz}.
\end{gathered}
\end{equation}

This system of equations can be written as a set of 2-forms on an
appropriate jet-bundle \cite{H97}. Following the techniques developed by
Estabrook and Wahlquist \cite{EW75}, a complete prolongation structure can
be computed \cite{F97}, in preparation for applying known solution
techniques.

Here we are interested in deriving one particular B{\"a}cklund transform. To
this end, it is useful to introduce new variables $c$ and $s$ defined by
\begin{equation}\label{xform}
	\chi_t - 2Kz = \frac{c}{c^2-s^2}\qquad
	\chi_z + 2Kt = \frac{s/a}{c^2-s^2}.
\end{equation}
The transformation from $(\chi_t,\chi_z)$ to $(c,s)$ is a smooth coordinate
transformation away from the origin $c = s = 0$. With these variables,
equations \eqref{mixed} become
\begin{equation}
\begin{aligned}
	(c^2 + s^2)(c_t - (as)_z) - 2sc(s_t - (ac)_z) &=0\\
	(c^2 + s^2)(s_t - (ac)_z) - 2sc(c_t - (as)_z) &=0.
\end{aligned}
\end{equation}
Since the origin $c=s=0$ is excluded, we conclude that Waylen's equation is
equivalent to the remarkably simple system
\begin{equation}
	s_t = (ac)_z \qquad c_t = (as)_z.
\end{equation}

Each of these equations can be expressed in potential form by introducing
potentials $x$ and $y$ such that
\begin{equation}\label{map}
\begin{aligned}
	x_t &= ac  \qquad  	&x_z &= s\\
	y_t &= as		&y_z &= c.
\end{aligned}
\end{equation}
Equations \eqref{map} (together with the transformation \eqref{xform})
constitute a B{\"a}cklund map between the variables $(a,\chi)$ and
$(x,y)$. Inverting the map \eqref{map} leads immediately to a first-order
equation
\begin{equation}\label{first}
	x_t x_z - y_t y_z = 0.
\end{equation}
Applying the integrability condition $\chi_{tz}=\chi_{zt}$ gives rise to
another equation of second order, which in turn has a first integral
\cite{H97}
\begin{equation}\label{integral}
	x_t y_z - x_z y_t = e^{4Ky} F(x)
\end{equation}
for an arbitrary function $F$ of one argument. The integrability conditions
on the B{\"a}cklund map \eqref{map} are now exhausted, so the pair of equations
\eqref{first}, \eqref{integral} constitutes a B{\"a}cklund transform for Waylen's
equation \eqref{mixed}.

To complete the link between Waylen's equation and the similarity variables
discussed in the previous section, we use the B{\"a}cklund map \eqref{map} and
transformation \eqref{xform}, to write
\begin{equation}
	\chi_t - 2Kz = \frac{y_z}{y_z{}^2 - x_z{}^2}\qquad
	\chi_z + 2Kt = \frac{y_t}{y_t{}^2 - x_t{}^2}.
\end{equation}
Making use of the relation \eqref{first}, these two equations can be
written
\begin{equation}
	\chi_t - 2Kz = \frac{x_t}{x_t y_z - x_z y_t}\qquad
	\chi_z + 2Kt = \frac{x_z}{x_t y_z - x_z y_t}.
\end{equation}
Relabelling
\begin{equation}\label{correspondence}
	x = U \qquad	y = \frac{1}{4K}\log V,
\end{equation}
these equations become
\begin{equation}
	\chi_t - 2Kz = \frac{4KV U_t}{U_t V_z - U_z V_t}\qquad
	\chi_z + 2Kt = \frac{4KV U_z}{U_t V_z - U_z V_t},
\end{equation}
which agrees completely with the invariance condition \eqref{simvar1}.

Using the correspondence \eqref{correspondence}, we can re-write the
first-order B{\"a}cklund transform equation \eqref{first} as
\begin{equation}\label{constraint}
	(4KV)^2 U_t U_z - V_t V_z = 0,
\end{equation}
while the first integral \eqref{integral} becomes
\begin{equation}\label{integral1}
	U_t V_z - U_z V_t = 4KV^2 F(U).
\end{equation}

The important point here is that \eqref{integral1} is a consequence of the
invariance conditions \eqref{simvar} as it can be derived directly using
$\chi_{tz} = \chi_{zt}$. The constraint \eqref{constraint} however cannot
be derived from the invariance conditions: it is an additional restriction
on the similarity variables. Recalling that this restriction arose from a
B{\"a}cklund transform of Waylen's equation \eqref{waylen}, we conclude
that the latter selects a restricted class of the possible similarity
variables. It follows that the exact forms \eqref{exactform} of the
invariants are not valid for all similarity solutions, but only for a
subset.

\section{Summary}

We have described Waylen's exact similarity solution results for an
axisymmetric vacuum space-time. Our analysis shows that Waylen's key
function $\chi$ and constant $K$ parameterise the infinite-dimensional Lie
pseudogroup of point symmetries for the vacuum field equations.
Furthermore, Waylen's equation \eqref{waylen} for the key function
represents a restriction on the class of similarity solutions which is
expressed via a B{\"a}cklund transform as a constraint \eqref{constraint}
on the similarity variables. Finally, we have shown that all similarity
solutions (whether Waylen's equation is satisfied or not) result in an
additional Killing vector (homothetic if $K\not=0$) for the space-time
metric.

\section*{Acknowledgements}
The determination of the point symmetries of the field equations
\eqref{main} was carried out using Langton's enhancements \cite{L97} to
Kersten's REDUCE package \cite{K87}. The authors are grateful to JD~Finley
and PC~Waylen for discussions, as well as G~Haager and H~Stephani for
suggesting the check for Killing vectors in section \ref{symmetries}.


\begin{thebibliography}{99}

\bibitem{W87}
PC Waylen, ``{\em Canonical solution of the equations of axisymmetric
gravitation including time dependence}'', Proc R Soc Lond A {\bf
411}(1987)49--57
\bibitem{W93}
PC Waylen, ``{\em Exact treatment of time-independent axisymmetric
gravitation}'', Proc R Soc Lond A {\bf 440}(1993)711--715
\bibitem{L97} BT Langton, ``{\em Lie Symmetry Techniques for Exact Interior
Solutions of the Einstein Field Equations for Axially Symmetric,
Stationary, Rigidly Rotating Perfect Fluids}'', PhD Thesis, University of
Sydney, 1997
\bibitem{K87} 
PHM Kersten, ``{\em Infinitesimal symmetries: a computational approach}'',
CWI Tract 34 (Centre for Mathematics and Computer Science, Amsterdam, 1987)
\bibitem{H97}
D Hartley, ``{\em A B{\"a}cklund transform for Waylen's equation}'', to appear
in ``{\em Proceedings of the 6th Monash General Relativity Workshop}''
ed AWC~Lun
\bibitem{EW75}
HD Wahlquist and FB Estabrook, ``{\em Prolongation structures of nonlinear
evolution equations}'', J Math Phys {\bf 16}(1975)1--7
\bibitem{F97}
JD Finley, private communication

\end{thebibliography}
\end{document}